\documentclass[fleqn,10pt]{wlscirep}
\usepackage[utf8]{inputenc}
\usepackage[T1]{fontenc}
\usepackage{xcolor}

\title{A dataset of direct observations of sea ice drift and waves in ice}

\author[1,*]{Jean Rabault}
\author[2,3]{Malte M\"uller}
\author[4]{Joey Voermans}
\author[5]{Dmitry Brazhnikov}
\author[6]{Ian Turnbull}
\author[7]{Aleksey Marchenko}
\author[8]{Martin Biuw}
\author[9]{Takehiko Nose}
\author[9,10]{Takuji Waseda}
\author[11]{Malin Johansson}
\author[12,13]{{\O}yvind Breivik}
\author[14]{Graig Sutherland}
\author[12]{Lars Robert Hole}
\author[5]{Mark Johnson}
\author[15]{Atle Jensen}
\author[15]{Olav Gundersen}
\author[16]{Yngve Kristoffersen}
\author[4]{Alexander Babanin}
\author[1, 17]{Paulina Tedesco}
\author[2, 3]{Kai Haakon Christensen}
\author[8]{Martin Kristiansen}
\author[12]{Gaute Hope}
\author[9]{Tsubasa Kodaira}
\author[11]{Victor de Aguiar}
\author[11]{Catherine Taelman}
\author[11]{Cornelius P Quigley}
\author[18]{Kirill Filchuk}
\author[19]{Andrew R Mahoney}

\affil[1]{Norwegian Meteorological Institute, IT Department, Oslo, 0313, Norway}
\affil[2]{Norwegian Meteorological Institute, R\&D Deparment, Oslo, 0313, Norway}
\affil[3]{University of Oslo, Deparment of Geosciences, Oslo, 0313, Norway}
\affil[4]{University of Melbourne, Department of Infrastructure Engineering, Melbourne, 3010, Australia}
\affil[5]{University of Alaska Fairbanks, College of Fisheries and Ocean Sciences, Fairbanks, 99775, USA}
\affil[6]{C-CORE, Captain Robert A. Bartlett Building, Morrissey Road, St. John's, Newfoundland and Labrador A1B 3X5, Canada}
\affil[7]{The University Centre in Svalbard, Arctic Technology Deparment, Longyearbyen, 156 N-9171, Norway}
\affil[8]{Institute of Marine Research, Fram Centre, P.O. Box 6606 Langnes, 9296 Tromsø, Norway}
\affil[9]{The University of Tokyo, Graduate School of Frontier Sciences, Kashiwa 277-8561, Japan}
\affil[10]{Japan Agency for Marine-Earth Science and Technology, Yokosuka 237-0061, Japan}
\affil[11]{UiT The Arctic University of Norway, Department of Physics and Technology, 9037 Tromsø, Norway}
\affil[12]{Norwegian Meteorological Institute, R\&D Department, Bergen, 5007, Norway}
\affil[13]{University of Bergen, Geophysical Institute, Bergen, 5007, Norway}
\affil[14]{Environment and Climate Change Canada, Environmental Numerical Prediction Research Department, Dorval, QC  K1A 0H3, Canada}
\affil[15]{University of Oslo, Department of Mathematics, Oslo, 0313, Norway}
\affil[16]{University of Bergen, Department of Earth Science, Bergen, 5007, Norway}
\affil[17]{University of Oslo, Department of Physics, Oslo, 0313, Norway}
\affil[18]{Arctic and Antarctic Research Institute (AARI), St. Petersburg, Russian Federation}
\affil[19]{University of Alaska Fairbanks, Geophysical Institute, Fairbanks, 99775, USA}

\affil[*]{corresponding author: Jean Rabault (jean.rblt@gmail.com)}

\begin{abstract}
Variability in sea ice conditions, combined with strong couplings to the atmosphere and
the ocean, lead to a broad range of complex sea ice dynamics. More in-situ measurements
are needed to better identify the phenomena and mechanisms that govern sea ice growth,
drift, and breakup. To this end, we have gathered a dataset of in-situ observations of sea ice drift
and waves in ice. A total of 15 deployments were
performed over a period of 5 years in both the Arctic and Antarctic, involving 72 instruments. These provide both GPS
drift tracks, and measurements of waves in ice. The data can, in turn, be used
for tuning sea ice drift models, investigating waves damping by sea ice, and helping
calibrate other sea ice measurement techniques, such as satellite based observations.
\end{abstract}
\begin{document}

\flushbottom
\maketitle

\thispagestyle{empty}

\section*{Background \& Summary}

Sea ice is a major component of the global Earth ecosystem: it covers around 7\% of the
global oceans, averaged over a year \cite{parkinson1997earth}, and strongly
modulates the coupling between the ocean and the atmosphere, as well as the global energy
balance of the polar regions \cite{johannessen2004arctic, vihma2014effects}. For example, a strong nonlinear coupling exists between
ocean conditions and sea ice extent in the Arctic, due to the effect of ocean waves
\cite{https://doi.org/10.1002/2014GL059983}. Indeed, as sea ice extent decreases in the
Arctic basin, a new area of open ocean emerges between the polar ice cap and the
surrounding continental landmasses. This, in turn, creates a new region of fetch where
waves can grow larger before interacting with the sea ice. As a consequence of these
larger waves, more sea ice gets broken which also accelerates melting, leading to a
positive feedback loop. Similar couplings exist due to, for example, the albedo differences
between open ocean and sea ice \cite{curry1995sea}. These changes are exacerbated by the polar
amplification of climate change \cite{holland2003polar, dai2019arctic}.

Unfortunately, accurately predicting sea ice dynamics under the influence of waves, winds,
and currents, remains a challenging task \cite{squire2007ocean, golden2020modeling,
ardhuin2020ice, https://doi.org/10.1029/2019JC015354,
https://doi.org/10.1002/2014GL059983, https://doi.org/10.1002/2015JC010881,
https://doi.org/10.1002/2017JC013275, squire2020ocean, smith2018modelling,
thomson2016emerging, marchenko2019wave, sutherland2022estimating, herman2018wave,herman2019wave, herman2019wave2, herman2021spectral, herman2018floe}. This is due to the diversity of both sea ice and
weather conditions that can be found in the polar regions, as well as the wide range of
physical mechanisms at stake, including sea ice breaking, melting, freezing, collisions
between ice floes, drifting, wave diffraction, reflection, viscoelastic effects in the
ice, and turbulence in the water under the ice \cite{zhao2018three, voermans2019wave,
rabault2019experiments, horvat2017evolution, sutherland2019two,
https://doi.org/10.1029/2019JC015354, roach2019advances, horvat2015prognostic,
montiel2017modelling, SREE2020102233, tc-2021-210, marchenko2017field,
sutherland2022estimating, sutherland2017attenuation, voermans2020experimental,
Li2021LaboratorySO, marchenko2021laboratory, doi:10.1063/5.0088953}. This makes the development of fully coupled
ocean-atmosphere-sea ice-waves models a challenging task \cite{williams2017wave,
batrak2018atmospheric, li2021effects, roach2018emergent}.

As a consequence, careful validation and calibration of sea ice models over a wide range of
sea ice conditions is critical. Unfortunately, calibration data about sea ice drift and
waves in ice are scarce, especially in the Marginal Ice Zone (MIZ), i.e. the sea ice area
closest to the open water, which is most influenced by waves from the open ocean \cite{williams2013wave}.
Gathering in-situ data in this region is made challenging due to the statistically short
lifetime of instrumentation under dynamic sea ice conditions, while remote sensing is, at
present, challenging to use in a systematic way, and not yet able to provide all the information needed to calibrate models, especially
regarding wave properties \cite{https://doi.org/10.1029/2022JC018654}.

\begin{figure}[ht]
\centering
\includegraphics[width=\linewidth]{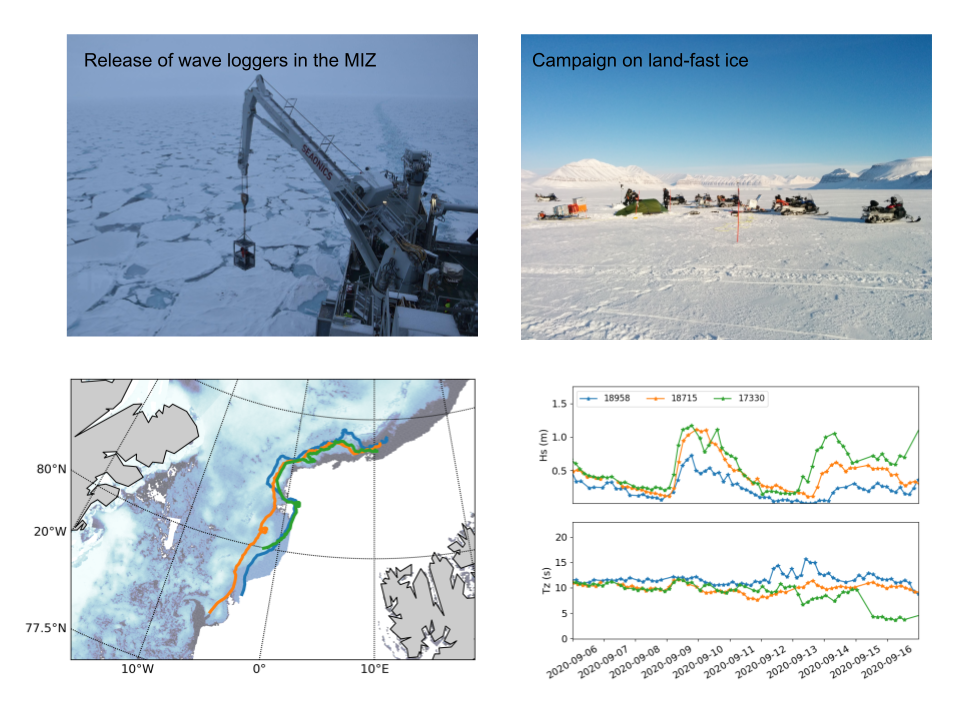}
\caption{Sea ice can come in many forms, from fields of small broken floes (top left picture), to continuous pack ice, to landfast ice (top right picture). This
has a strong, complex influence on wave damping and sea ice drift. To illustrate the effect of sea ice on wave damping, we show the map presenting Sea Ice Concentration (SIC, bottom left) around
3 drifters corresponding to the 2020-07 deployment on 2020-09-16, and the corresponding drifter trajectories. The drifters are at different depth into the Marginal Ice Zone (MIZ), which implies different
degrees of frequency-dependent significant wave heigth damping and associated peak frequency shift (bottom right). The field data released in this manuscript
features many similar events, that can be used to tune numerical models.}
\label{fig:main_illustration}
\end{figure}

To help close the gap in sea ice data in the MIZ and other regions affected by waves
propagating from the open ocean, we have focused on deploying instruments in both the Arctic and Antarctic. These instruments collect both
drift information and wave energy spectrum. Two situations are of particular interest.
The first corresponds to sea ice in the MIZ, where broken ice floes drift and break up under the influence
of currents, winds, and waves. In these conditions, we usually deploy several instruments perpendicular to the MIZ
edge, going successively deeper into the ice, so that we can quantify the effect of sea ice concentration
on drift patterns, as well as wave damping into the MIZ, as illustrated in Fig. \ref{fig:main_illustration}. The second canonical situation of interest is
continuous pack ice or landfast ice just outside of the MIZ. Remotely forced waves that are only partially damped in the MIZ ca nreach this region. This, in turn,
can lead to breakup of the continuous ice, which can drift away towards the open ocean where it
breaks down further and melts. In these conditions, we deploy instruments close to the limit between the MIZ
and the continuous ice, in order to study the relationship between incoming wave energy and continuous ice
breakup.

In the present dataset, we release
data collected over the last 5 years by 13 groups, which corresponds to 15
deployments, involving a total of 72 instruments, covering both the Norwegian, Russian, East Greenland, and Alaskan Arctic,
and several regions in the Antarctic. These data can, in turn, be used for investigating the effect of
winds, currents, and waves on sea ice, to both identify physical processes, and help further develop and calibrate fully coupled models, in particular, regarding
sea ice drift, waves in ice, and sea ice breakup. Moreover, these
in-situ data are ground observations necessary to help calibrate satellite derived algorithms
used in remote sensing of the sea ice.

\section*{Methods}

In the course of the data collection, several instrument models have been used. These are,
the instrument "v2018" \cite{rabault2020open}, the instrument "v2021" \cite{rabault2022openmetbuoy, 9754500},
the Sofar Spotter buoy \cite{PerformanceCharacteristicsofSpotteraNewlyDevelopedRealTimeWaveMeasurementBuoy},
commercial Global Positioning System (GPS) drifters with Iridium communication ability, and the Ice Wave Rider (IWR) 
\cite{johnson2020measuring}.
All instruments use GPS to measure geographical location. The instrument v2018, v2021, and IWR,
use acceleration measurements from inertial measurement units to measure wave motion. By contrast,
the Sofar Spotter uses GPS to measure the wave motion.

In the
following, we describe how measurements of drift and waves are performed, and we pinpoint
differences between the instruments when relevant. The methodology used by the Sofar Spotter
buoy is only partially known, as this is a close source, black box instrument. By contrast,
the firmware and post processing code for the v2018, v2021 are fully open source, while the IWR is simply performing logging
of an inertial measurement unit, with known configuration \cite{johnson2020measuring}. Therefore,
and in addition to the detailed self-contained methodology description we provide here, the open source codes (available Github at
\cite{code_v2018} and \cite{code_v2021},
respectively) can also be used as a source for technical details around the methodology used
by the instruments v2018 and v2021.

\subsection*{Drift measurements using the Global Positioning System}

Drift tracks are measured using a GPS receiver. No processing
of any form is applied on the output produced by the GPS module. The GPS module has
built-in hardware and software that ensure that only valid GPS positions are produced.

\begin{itemize}
    \item In the case of the v2018, the full GPS National Marine Electronics Association
    (NMEA) output string is transmitted through Iridium. In addition, all the data collected by the instrument v2018 are stored on
    an internal SD card, so that, if the instrument v2018 is collected
    at the end of the deployment, the same GPS data are available there. The GPS measurements are performed with a period of approximately 3 hours.
    \item In the case of the v2021, only the UTC date and time as well as the latitude and
    longitude are transmitted. Other information is not transmitted, to save memory and
    cut satellite transmission costs. However, a higher GPS position sampling rate is
    applied, typically every 30 minutes.
    \item Similarly, the Sofar Spotter transmits GPS UTC time, latitude and longitude as a part of its iridium
    communications.
    \item The IWRs log GPS UTC time and geographic coordinates hourly to the internal SD 
    card, without performing satellite transmission. Such routine is possible since IWRs deployments are either in 
    ice camps or on shore-fast ice, and the data are retrieved together with the instruments at the end of the deployments.
    \item Commercial drifters also transmit GPS UTC time, latitude, and longitude.
\end{itemize}

\subsection*{Wave measurements using GPS}

The Sofar Spotter uses GPS data in order to directly measure 2D wave surface displacement at 2.5Hz in the North-South and East-West directions,
from which to compute the wave properties assuming that the underlying signal corresponds to
the circular wave orbital velocity \cite{PerformanceCharacteristicsofSpotteraNewlyDevelopedRealTimeWaveMeasurementBuoy}. According to the datasheet delivered by Sofar, the typical
resolution for significant wave height measurements is +/-2cm, depending on the conditions and sky view. The details of the implementation regarding how the raw data are filtered and how the
wave spectrum is computed are, as far as the authors know, not available for this commercial close source instrument.
The produced outputs include both wave statistics (significant wave height, and various wave period estimates), full
wave spectra, and directional information (some of these outputs can be switched on and off individually of each other).
The Sofar Spotter is considered a well-tested instrument, with several thousands units deployed according to the manufacturer,
and it has been used in a number of peer reviewed measurement campaigns in the sea ice \cite{voermans2020experimental, KODAIRA2021100567, nederhoff2022effect, kousal2022two}.

\subsection*{Wave measurements using Inertial Measurement Unit data}

The instruments v2018, v2021, and IWR, use Inertial Measurement Units (IMUs) to measure the
wave motion. In the following paragraphs, we outline
the main lines of the data acquisition and the wave processing algorithms, though the reader
curious of the exact, in-depth technical details regarding the open source instruments,
is referred to the technical papers
\cite{rabault2020open, rabault2022openmetbuoy, johnson2020measuring} that go deeper into the exact implementation, or to the code implementing the data
processing (see the previous Github links).

\subsubsection*{Measurement of the wave vertical acceleration}

The first step in measuring waves is to record the wave signal, i.e acceleration due to the wave
motion. A duration of 20 minutes
is used as the default time segment used to produce one wave spectrum and its associated
statistics, and we used a wave acceleration sample frequency of 10Hz unless specified otherwise. In order to
gather this wave signal:

\begin{itemize}
    \item In the case of the v2018 \cite{rabault2020open}, all the data collection and Kalman filtering process
    is implemented by the thermally calibrated Vectornav VN100 Inertial Measurement Unit (IMU)
    that is used in the instrument \cite{rabault2016measurements, rabault2017measurements, sutherland2016observations}. The VN100 is an industry-grade IMU, and the details of
    the Kalman filter and signal processing algorithms running on it are proprietary and close source. The VN100 IMU
    outputs a number of variables, including accelerations measured in the Earth frame of
    reference (North, East, Down directions, referred further as NED). The Down (D)
    component of the acceleration, $acc_D$, is obtained from the VN100 at a frequency of
    10Hz and used for evaluating waves statistics. This 10Hz value is obtained by
    applying a Kalman filter to  the raw sensor output (that is obtained at several kHZ), which translates into a
    800Hz processed output. After that, a running
    average over 80 samples is performed to obtain the final 10Hz value. In the
    case when instruments are collected at the end of deployment, the full time series of
    the VN100 output at 10Hz, is available on the internal SD card. This includes both raw
    sensor measurements, Kalman filter state, heading and orientation output, and acceleration
    in the Earth frame of reference.
    
    \item In the case of the v2021 \cite{rabault2022openmetbuoy}, the temperature-compensated 9-degree-of-freedom (9dof) sensor
    measures the accelerations, angular rates, and the local magnetic field, each over
    3 axis (the X, Y, and Z axis of the 9dof sensor, in its own frame of reference
    attached to the microchip itself). These raw measurements are performed at 800Hz.
    Averaging and n-sigma filtering over 8 raw samples is used to downsample the signal from 800Hz to 100Hz,
    reject possible outliers, and reduce noise. An open
    source Kalman filter implementation is then used to fuse the data at 100Hz update
    frequency. The Kalman filter produces a unit quaternion estimate $q$ of the
    absolute orientation of the sensor relative to the Earth frame of reference. This
    information, combined with the acceleration measurements $acc_{refSensor} = [acc_X,
    acc_Y, acc_Z]$ in the sensor frame of reference, allows to retrieve the acceleration
    of the sensor in the Earth frame of reference $acc_{refEarth} = [acc_N, acc_E,
    acc_D]$, by applying the rotation described by the quaternion $q$ on $acc_{refSensor}$.
    At present, magnetometer calibration is not performed carefully enough to
    trust directional information relative to the magnetic North direction (this is
    ongoing work for further deployments). Therefore, the only data used at the moment
    are the vertical accelerations. These are averaged from the 100Hz $acc_D$ output
    into a 10Hz wave acceleration signal.
    
    \item  Similar to the v2018 instrument, the IWR uses the Vectornav VN100 IMU. Processing and settings similar to the ones
    used with the instrument "v2018" are used by the IWR. However, the IWR records
    wave acceleration continuously, and stores it on-the-fly on an attached SD card, so that the time series
    of the wave displacement are available. Initial deployments 
    were storing every 80$^{th}$ value at the rate of 10 Hz. Later, to improve data quality, the 
    sensors were reprogrammed to average sequences of 80 values, which were then saved at 10 Hz. 
    Data output differed between the deployments, but always included $[acc_N, acc_E,
    acc_D]$ along with yaw pitch roll angles of the IWR. The accelerations were output with 
    constant vertical gravity acceleration removed.
\end{itemize}

At this stage, the vertical acceleration, which is the superposition of the gravity and
wave acceleration (except for the IWR, where gravity is removed), is available at 10Hz for further processing in the case of either the
v2018 or the v2021 (the IWR is a pure logger that is not equipped with iridium transmissions
and does not perform in-situ processing of the data).

\subsubsection*{Estimation of the wave spectrum}

The 10Hz wave acceleration component alongside the Down (D) direction in the Earth frame
of reference is then used to compute the wave spectrum and its statistics. For this, two
variants of the same core methodology are applied in the case of the instrument v2018 and
v2021.

\begin{itemize}
    \item In the case of the instrument v2018, details of the methodology are presented in the technical
    paper describing the instrument \cite{rabault2020open}. We reproduce the main lines of the processing here. First, the vertical
    acceleration is integrated twice in time using the methodology of previously developed instruments \cite{kohout2015device},
    which is done in Fourier space by using the frequency response weights of $1/
    \omega^2$ and a half-cosine taper for the lower frequencies to avoid an abrupt cut-off
    \cite{tucker2001waves}, corresponding to:

\begin{equation}
    \eta(t) = Re(IFFT[H(f)FFT(acc_D)]),
\end{equation}

\noindent where $Re$ indicates the real part of the signal, IFFT stands for the Inverse
Fast Fourier Transform, FFT the Fast Fourier Transform, and $H(f)$ is the half-cosine taper
function:

\begin{equation}
    H(f) =
    	\begin{cases}
		0, & 0 < f < f_1 \\
		\frac{1}{2}\left[ 1 - \cos{ \left(\pi\frac{f-f_1}{f_2-f_1}\right)\left(\frac{-1}{2\pi f^2}\right) }\right], & f_1 \le f \le f_2 \\
		\frac{-1}{2\pi f^2}, & f_2 < f < f_c, 
	\end{cases}
\end{equation}

\noindent where $f$ is the frequency,  $f_c$ is the Nyquist frequency, and $f_1 = 0.02
\mathrm{Hz}$ and $f_2 = 0.03 \mathrm{Hz}$ are the half-cosine taper corner frequencies,
similar to \cite{kohout2015device}.

Following the calculation of the time series for the wave elevation $\eta(t)$ at 10Hz, the
wave elevation spectrum $S(f)$ is estimated using the Welch method on 12000 samples (20
minutes at 10Hz), using a Hanning window of length 1024 samples per segment, and 50\%
overlap. In addition, the spectral moments $m_0$, $m_1$, $m_2$, $m_4$ of the wave spectrum
are computed, following:

\begin{equation}
    m_n = \int_{0.05}^{0.25} f^n S(f) df.
\end{equation}

This allows to compute the usual wave statistics, i.e. the significant wave height
calculated from the time series, $H_{St} = 4 std(\eta)$, the significant wave height
calculated from the spectral moment $H_{S0} = 4 \sqrt{m_0}$, the spectra peak period $T_p$
corresponding to the maximum of the spectrum, the zero-upcrossing period $T_z =
\sqrt{m_2/m_0}$, and the average crest period $T_c = \sqrt{m_2/m_4}$.

In addition to these data, some directional spread estimates are computed, though these
are not as carefully validated and their accuracy may be lower. The reader who
wants to use these is invited to read about the technical details in the technical paper \cite{rabault2020open}, and to
implement their own quality checks if they want to use directional information.

  \item In the case of the instrument v2021 \cite{rabault2022openmetbuoy}, a slightly simpler methodology is used. The
  Power Spectral Density (PSD) for the vertical wave acceleration
  $PSD_{accD}$ is directly calculated from the vertical wave acceleration data at 10Hz, by
  applying the Welch method on 20.48 minutes of data (so that the exact number of samples
  is a multiple of $2^{11}$, which makes the calculation of the FFTs and the splitting
  into segments simpler and faster). Each segment for the Welch method has a length of
  2048 samples, 75\% overlap is used between the segments, and an energy-preserving Hanning window is
  used.

From there, the spectrum for the wave vertical elevation $S(f)$ can be retrieved as:

\begin{equation}
    S(f) = \frac{PSD_{accD}(f)}{(2 \pi f)^4}.
    \label{psd_to_spectrum}
\end{equation}

At this stage, the spectral moments, as well as $Hs$, $T_z$, and $T_c$, can be calculated
in the same way as for the instrument v2018.

\end{itemize}

\subsection*{Data transmission and decoding}

Data are transmitted from the sea ice by the instruments v2018, v2021, and the Sofar Spotter,
using the Iridium communication network and the
Short Burst Data (SBD) protocol, which allows messages of size up to 340 bytes.

\begin{itemize}
    \item In the case of the instrument v2018, the wave spectrum $S(F)$ is downsampled
    into 25 logarithmically equally spaced bins between 0.05Hz and 0.25Hz, and transmitted
    together with the wave statistical parameters. 
    
    \item In the case of the instrument v2021, the full PSD of the wave vertical
    acceleration $PSD_{accD}(f)$ between 0.05Hz and 0.3Hz is transmitted, alongside with
    the wave statistical parameters, though these are merely a cross validation of the
    spectrum, in the sense that they can be derived directly from the transmitted full
    spectrum. A simple script, which performs no further processing
    of the data, is then used to decode the binary SBD messages and apply the translation
    from $PSD_{accD}(f)$ to $S(f)$ following Eqn. \ref{psd_to_spectrum}.
    
    \item In the case of the Sofar Spotter, a variety of modes are available, in which either
    the full spectrum, or only integrated quantities, are transmitted. The data are available as
    JSON files, which are provided directly by Sofar by unpacking the binary data.
\end{itemize}

Commercial iridium trackers provide the track information data through a web-based API that abstract
the details of the data transmission and processing protocol.

\section*{Data Records}

All our high level, processed data are provided as netCDF4-CF files on the THREDDS data server of the Norwegian
Meteorological Institute through the Arctic Data Center (ADC) database, at the following address, and in the folder structure therein:
\url{https://adc.met.no/datasets/10.21343/azky-0x44} \cite{thredds_data}.
Extensive metadata are provided, so that the different data fields are self documenting, following the FAIR data principle \cite{wilkinson2016fair}.
In addition, all the raw iridium data transmissions (hex-string binary data), raw SD card
files (binary data, available only when the instruments were recovered),
and scripts for reading these raw data files and processing them
into netCDF files, as well as example scripts of how the netCDF datasets can be read, are
available on Github in the folder structure of the following repository:
\url{https://github.com/jerabaul29/data_release_sea_ice_drift_waves_in_ice_marginal_ice_zone_2022} \cite{main_gh_repo}.

In the following, we provide a detailed overview of the different deployments, what kinds of instruments
and how many of them were used in each of these, and some background information about the sea ice conditions
and the data of interest. The time series obtained are of variable duration and some of them have holes in the
data. This is due, to the best of our knowledge, to the harsh conditions found on sea ice, rather than technical
issues with the instruments. Factors such as heavy snowfalls, polar bear destroying equipment, sea ice breakup,
ridging and rafting, are all susceptible to interrupt iridium communications or destroy prematurely the instruments
altogether.

The deployments in the Arctic are, in chronological order:

\begin{itemize}
    \item 2017-04: Arctic deployment in the MIZ that contains drift data in the Barents Sea. 8 GPS
    ice trackers were deployed on 6 ice floes in the Barents Sea South of Svalbard. The trackers
    send GPS data every 30 minutes, and were deployed during April 24-26 2017 from the Research Vessel
    Polarsyssel. Trackers 4610 and 8650 were deployed on the same ice floe, and trackers 5630 and 2470
    were deployed on the same ice floe, different from the previous one. The other trackers were deployed
    each on their respective ice floe. An overview of the deployment time for the different trackers is provided
    in Table \ref{tab:ice_tracker_deployment_2017}. The trackers could float and, therefore, the internal temperature
    records were used to determine the date and time when they entered water, as the temperature
    transitioned from a larger diurnal cycle when exposed to air on top of the ice, to a significantly
    more muted diurnal temperature variation when floating in open water. An analysis of the trackers’
    internal temperature records and the Svalbard regional daily ice maps from the Norwegian Meteorological
    Institute Ice Service led to the determination that the beacons most likely began falling into open water
    sometime on May 2, 2017 \cite{TURNBULL2022103463}. The dispersion of the ice floes was dominated
    by strong shearing within the local ice pack, coinciding with a rapid increase in the speeds of the
    local tidal currents, which was soon followed by a rapid increase in the wave energy. Each of the
    six tracked ice floes increased their observed drift speeds in sync with the increase in the local
    tidal current speeds at different times for each floe, but at approximately the same decrease in
    water depth as they reached the northern edge of Spitsbergen Bank. The rapid increase in the tidal
    currents was linked to the topographic enhancement of tidal motion near Hopen Island in the shallower
    waters of Spitsbergen Bank. The last transmission from the trackers was received 2017-07-15.

    \item 2018-03a: Arctic deployment in the MIZ, in the East Greenland Sea. The primary aim of the cruise
    was to monitor the production of seal pups. For that, 5 GPS and iridium trackers were deployed on
    large ice floes in the dense MIZ and drifter Southwestwards, following the East Greenland Current.
    GPS data sampling rate was 30 minutes. The trackers were deployed around March 20th, 2018, and
    the last tracker stopped transmitting on April 25th, 2018, though communications were unreliable after
    April 6th, 2018, possibly indicating that the trackers were covered by snow and ice. The trackers were
    not equipped with floatability equipment, so that the trackers are guaranteed to be on an ice floe
    for all the trajectory duration. More information is available in \cite{biuw2018report, biuw2022recent}.

    \item 2018-03b: Arctic deployment on pack ice in the Beaufort Sea during ICEX2018, U.S. Navy 
	 exercises. Two IWRs were deployed for 4 days near an ice camp on 2018-03-17. One instrument was located on 
	 level ice near the air strip. It recorded numerous events characterized by strong, 
	 high-frequency oscillations  produced by planes landing and taking off. The events matched the 
	 camp's logbook. The second instrument was placed in an ice rubble field nearby. It also 
	 measured several cases of strong accelerations, not coinciding with the first instrument 
	 observations. Their origin is unclear. The instruments were recovered on 2018-03-21. More information
	 are available in \cite{johnson2021observing}.

    \item 2018-04: Arctic deployment in the MIZ and drift ice in the Barents Sea. An ice tracker
    by Oceanetic Measurements Ltd was deployed on an ice floe to monitor the floe drift. GPS location
    was sent with an interval of 10 minutes. The tracker was deployed on April 27th, 2018 and transmitted
    until February 27th, 2019. The tracker drifted in the region of the Spitsbergen Bank for approximately
    6 months, though it was on an ice floe only until around May 3rd, 2018 according to ice charts from
    cryo.met.no , and after this date, the tracker was floating in ice-free waters. More information are
    available in \cite{marchenko2021properties}.

    \item 2018-09: Arctic deployment in the MIZ in the Barents Sea, in the context of the
    Nansen Legacy project, Physical Process Cruise 2018. In this deployment, a total of 4
    instruments v2018 were deployed while the icebreaker R/V Kronprins Haakon was
    traveling into the ice. The first instrument was deployed on a lone ice floe in the
    outer MIZ, while sea ice concentration (SIC) was about 1/10. The second instrument was
    deployed on an ice floe in dense drift ice, SIC about 5/10. The third instrument was
    deployed  at the very start of the close pack ice, SIC about 9/10. The fourth
    instrument was deployed in the close pack ice, SIC about 10/10. All instruments were
    deployed on 2018-09-19. The cruise report is available if more details are needed
    \cite{fer2020physical}. The instruments worked for around 2 weeks. However, a strong
    storm took place around 2018-09-24, and after this time, the spectra reported by the
    v2018 are very noisy with a lot of high frequency wave energy. This indicates that the
    supporting ice floes got broken into small pieces and started to drift in very open
    water, which is confirmed by SIC from models (see, for example, the discussions in
    manuscripts using the data \cite{sutherland2022estimating}). Therefore, the GPS drift tracks can be
    trusted over the whole duration of the dataset, but spectra dominated by high
    frequency signal correspond to small, broken ice floes drifting in very open water and
    experiencing erratic oscillations,
    rather than more classical waves in ice.
    
    \item 2020-03a: Arctic deployment on landfast ice in Gr\o{}nfjorden, Svalbard. Three v2018
    were deployed along the main axis of the fjord on continuous landfast ice
    between 10 and 13 March 2020. The first instrument was deployed at a distance of
    approximately 500 m from the unbroken ice edge, the second and third were
    deployed 600 and 700 m apart. At the time of deployment, the unbroken sea ice covered
    approximately half the length of the fjord. The other half was covered by broken ice
    and extended up to the entrance of the fjord. Ice thickness was measured at various
    locations near the instruments between 30 and 40 cm. Instruments were recovered after
    approximately two weeks on the 28th of March, 2020. At the time of retrieval all
    instruments were still deployed on unbroken ice and recording data. These data were used in
    some previous works \cite{voermans2020experimental, voermans2021wave}, and some additional information are available there.

    \item 2020-03b: Arctic deployment on pack ice in the Beaufort Sea during ICEX2020. The 
	deployment strategy was similar to the ICEX2018 campaign (see 2018-04). Deployment took place on 2020-03-08,
	and the instruments were collected on 2020-03-18. Similarly, the instrument close to 
	the air strip measured short-living, high-frequency events. No logging of arriving and leaving 
	planes was made. Both IWRs recorded two events of propagating flexural-gravity waves presumably 
	forced by strong winds during passing storms. More can information be found in \cite{johnson2021observing}.

    \item 2020-07: Arctic deployment in the MIZ in the Barents Sea over the Yermak Plateau,
    in the context of the Fram-2020 hovercraft expedition. The cruise report is available at
    \cite{yngve_report_2020} In this deployment, a total of
    6 instruments v2018 were deployed over the summer 2020 on drift ice, SIC ranging from
    approximately 3/10 to 10/10. One instrument was deployed
    2020-07-15, and transmitted until 2020-07-31. One instrument was deployed 2020-07-21,
    and transmitted until 2020-09-03. However, the energy content in the high frequency
    part of the spectrum of this instrument indicates that it likely drifted on a small,
    isolated, broken ice floe from around 2020-08-19. One instrument was deployed
    2020-08-14, and transmitted until 2020-09-08. Finally, 3 instruments were deployed
    around 2020-08-26, and transmitted until around 2020-09-24. Several wave events were
    consistently observed across the array of instruments deployed simultaneously, and
    clear wave damping is visible.
    
    \item 2021-02: Arctic deployment in the MIZ in the Barents Sea, East of Svalbard, in
    the context of the Nansen Legacy project, PC-2 Winter Process Cruise 2021. In this
    deployment, a total of 6 instruments v2018, and 11 prototypes of the instrument v2021 (of which 6 were
    equipped with wave measurements) were deployed from the icebreaker R/V
    Kronprins Haakon on its way up and down the East coast of Svalbard, in close drift ice. The details of the deployment
    are reported in the cruise report \cite{nilsen2021pc}. 3 instruments v2018 were deployed
    on the way into the ice, for SIC increasing from around 5/10, 9/10
    and 10/10. All the other instruments were deployed in close pack ice, for SIC
    of 10/10. The instruments gradually stopped transmitting as ice broke up, with the last transmission
    on ice taking place in late April.
    
    \item 2021-03: Arctic deployment on broken pack ice in the Beaufort Sea to accompany 
	measurements during Sea Ice Dynamic Experiment (SIDEx2021). The instruments were deployed on
	2021-03-06. In total six IWR instruments were 
	installed surrounding the ice camp, but only three were recovered. Due to battery 
	problems, the amount of collected data varied. The shortest observational interval was 2 weeks and 
    the longest one and a half month (it was collected on 2021-04-22). More information is available in \cite{johnson2021observing}.
	
	\item 2021-04: Arctic deployment on landfast ice near Utqiagvik, Alaska as a part of Integrated 
	System for Operations in Polar Seas project. Six buoys were placed in different locations on 
	thick landfast sea ice, and deployment took place on the 2021-05-07. Ice conditions varied between jumble ice and level ice of a refrozen 
	lead. Some instruments were located closer to the landfast ice edge. The measurements did not 
	exhibit any clear sign of waves. However, a wave event was detected via ground-based radar 
	interferometry. Respective vertical accelerations above IMU's noise level were also found 
	in the IWRs. The instruments were collected on the 2021-06-02. More on the deployment and results can be found in \cite{dammann2022observing}.
    
    \item 2021-09: Arctic deployment in the MIZ in the Laptev Sea, in the context of the NABOS campaign. Two buoys,
    one Sofar Spotter and one instrument v2021 packed in a Zeni floating enclosure, were deployed adjacent to the ice edge
    on 2021-09-15.
    The Sofar Spotter battery life at these latitudes and without solar input is only around 10 days, but by contrast,
    the v2021 instrument functioned for over 2 weeks. Several swell events were measured during the deployment,
    and the distance between the buoys allows to observe clear attenuation of
    the incoming swells by sea ice. The data are used in \cite{preprint_tak_et_al_laptev}, and the reader is referred to the corresponding
    work for further details about the deployment and the data collected. The last transmission included in
    the dataset took place on 2021-09-29.
    
    \item 2022-03: Arctic deployment in the MIZ in the East Greenland Sea, in the context
    of the seal pup monitoring cruise 2022. In this deployment, 2 instruments v2021 were installed on 2
    separate, neighboring
    medium size drifting ice floes (typical floe size: 25m; typical floe thickness: 1.5m) from
    the icebreaker R/V Kronprins Haakon on March 27th, 2022. In addition, 5 commercial gps and iridium drifters
    that only measured sea ice drift were deployed on large ice floes in the dense MIZ. The instruments drifted following the
    East Greenland current. The trajectories of the 2 instruments v2021 remained close (less than
    typically 2km apart) for around 2 weeks, before drifting slightly further apart from each other
    but heading in the same general direction. As a consequence, tracks and wave spectra are initially very similar,
    before clear differences in the wave spectra due to being at different depths in the MIZ get visible
    after around 2 weeks of drift. One of the instrument started to transmit data unreliably after
    around 4 weeks of activity, likely due to being covered by a layer of snow and ice that blocked iridium
    communications (since occasional transmissions were still recorded after messages started to come in
    unreliably, which did not indicate any technical issue with the instrument other than bad communications).
    The second instrument worked uninterrupted for over 4.5 weeks. The trajectories of the 5 commercial gps and iridium
    drifters remained close for most of the deployment. The last transmission was received on 2022-05-22. More information is available in the cruise report
    \cite{Biuw_Oigard_Nilssen_Stenson_Lindblom_Poltermann_Kristiansen_Haug_2022}.

\end{itemize}

The deployments in the Antarctic are, in chronological order:

\begin{itemize}
    \item 2020-01: Antarctic deployment on landfast ice on the eastern rim of the Amery
    Ice Shelf. Four instruments, two v2018 and two Sofar Spotter, were deployed along a transect perpendicular to
    the unbroken ice edge on 7 December 2019. The first Spotter was deployed 100-200 m
    from the unbroken ice edge, while a v2018 and the second Sofar Spotter (40 m apart from each other)
    were deployed at a distance of about 3.7
    km from the edge, and one more v2018 was deployed about 9.3 km from the unbroken ice edge. At the
    time of deployment, the ice was measured between 1 and 1.2 m thick. No wave events or
    drift was recorded until the 2nd of January 2020 when a large section of the fast ice
    broke and started drifting northward. The first v2018 stopped transmitting on 22
    January 2020, while the two Sofar Spotters stopped transmitting on 1 February 2020 (although
    one reconnected for half a day on 3 March 2020). The last message successfully
    transmitted by the final v2018 was on 10 March 2020. These data were used in papers
    \cite{voermans2020experimental, kousal2022two}, and some additional information are available there.
    
    \item 2020-11: Antarctic deployment on landfast ice north of Casey Station. Deployment
    consisted of two v2018, deployed 1.9 km apart in October
    2020. Instruments were recovered after about 3-4 weeks. Ice thickness was measured 1.1
    m thick during the deployment and 1.3 m during retrieval. As the deployment site was
    separated from the Southern Ocean open water by roughly 300 km of broken sea ice, only
    limited wave energy was observed. The sea ice at the deployment site remained unbroken
    throughout the deployment. More information is available in \cite{tc-2021-210}.
\end{itemize}

\begin{figure}[ht]
\centering
\includegraphics[width=\linewidth]{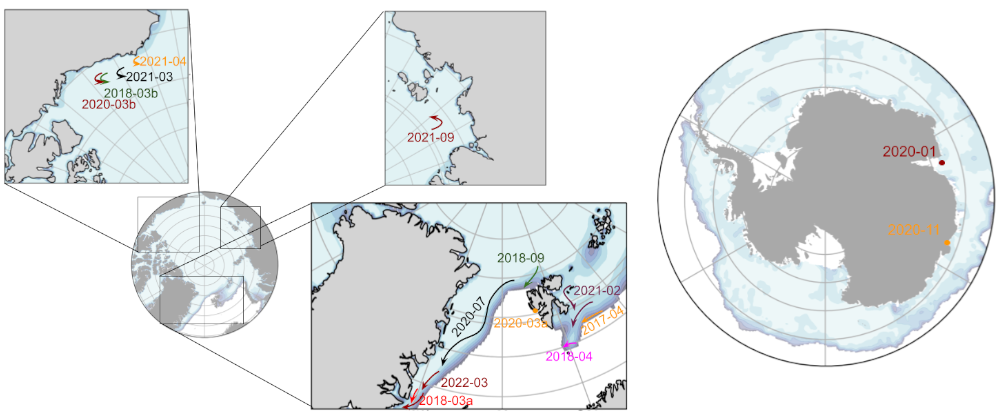}
\caption{Overview of the deployments present in the dataset. The SIC map show the averaged SIC over the local winter month in the Arctic and Antarctic.}
\label{fig:overview_deployments}
\end{figure}

A summary of the deployments is provided in Table \ref{tab:deployments}, and in Fig. \ref{fig:overview_deployments}.

\section*{Technical Validation}

There is no need for validation regarding GPS position measurements, as these are well established
sensors with a well known accuracy (typically +-5m).

Regarding wave data, we are only using well established, longstanding methodologies. The sensors used are
thermally calibrated over a range that typically exceeds the range of conditions found in the field in the MIZ (the Vectornav VN100 used in the v2018,
is thermally calibrated over the full temperature range from -40 to +85 C, while the ST-Microelectronics ISM330DHCX used in the v2021 is calibrated from -40 to
+105 C, according to their respective datasheets). Therefore, we consider that the data acquisition for the wave acceleration by itself
does not need additional validation, and users should refer to the datasheets of the corresponding sensors for further
information. As a side note, validation of the accelerometer data was performed, either directly (see Fig. 1 of \cite{rabault2016measurements} for the
test and validation performed for the VN100), or indirectly (by validating the accuracy of wave spectra, see next paragraph). In practice, however,
raw data from the instrument are only available for a couple of deployment of the instruments v2018 (when the SD card could be recovered),
and for the IWR deployments, for which the data provided are always the timeseries of the IMU output.

\begin{figure}[ht]
\centering
\includegraphics[width=0.90\linewidth]{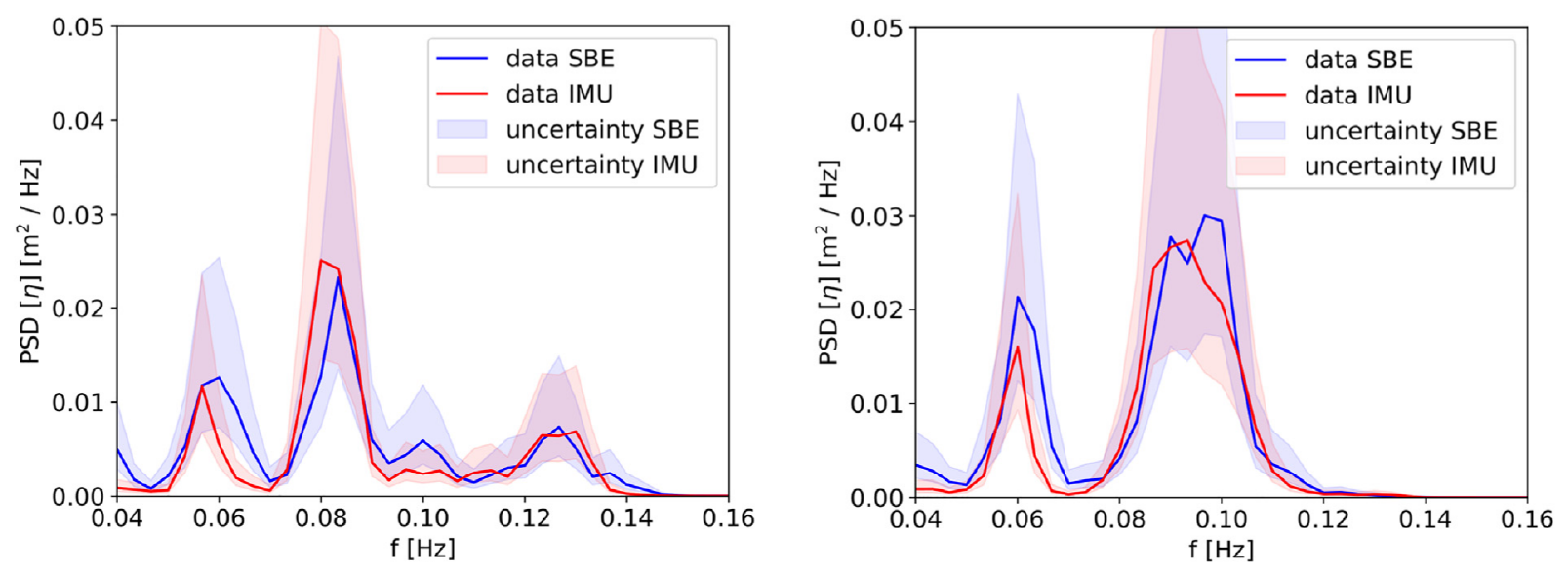}
\caption{Validation of the instrument v2018 IMU measurement and processing (IMU curves) against co-located pressure-sensor-based
measurements of waves using a Seabird pressure sensor (SBE curves), reproduced from \cite{rabault2020open}. Shaded areas indicate the 5-sigma confidence
intervals. Agreement within the confidence intervals is observed.}
\label{fig:v2018_validation_seabird}
\end{figure}

\begin{figure}[ht]
\centering
\includegraphics[width=0.60\linewidth]{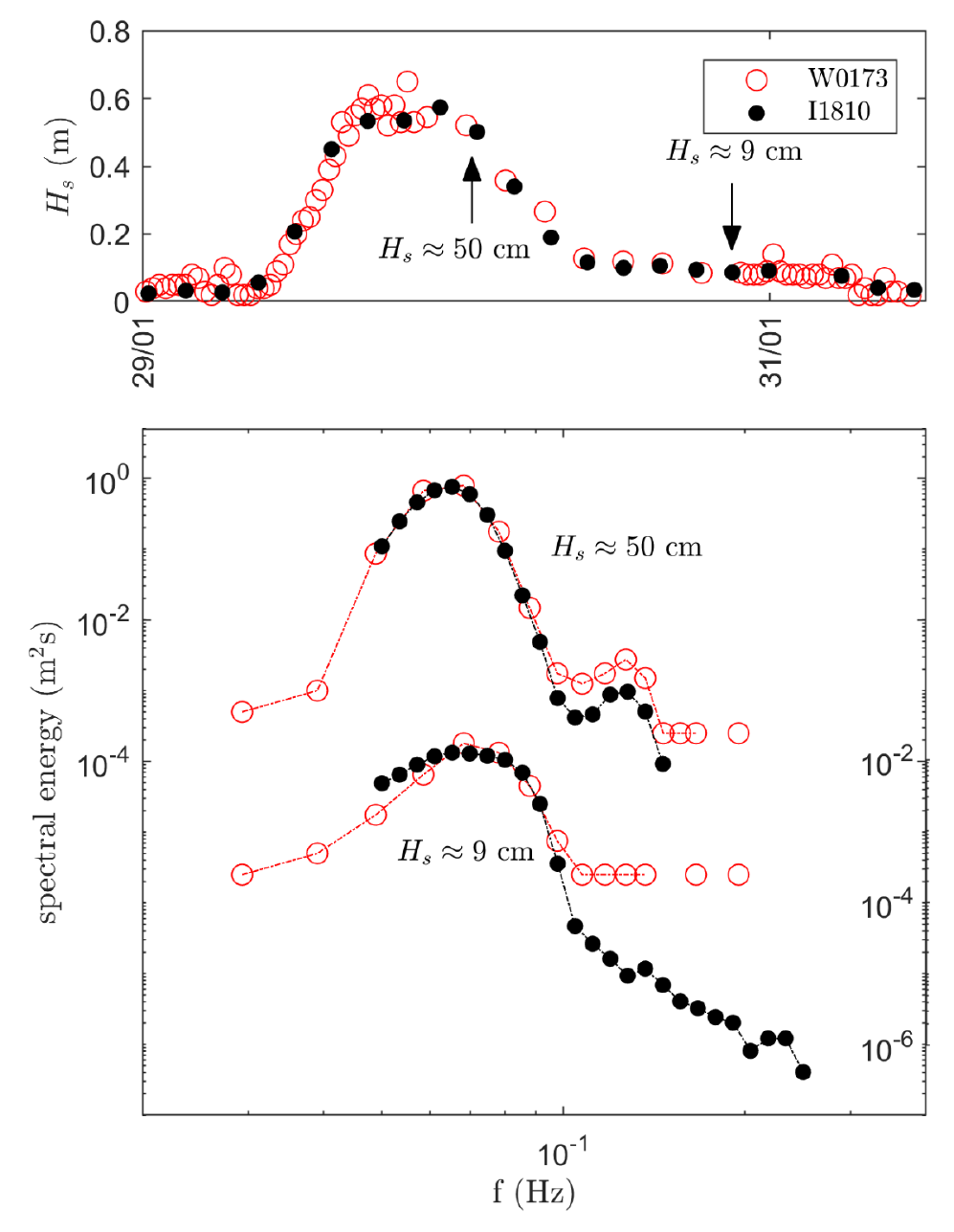}
\caption{Validation of the instrument v2018 (I810) against the Sofar Spotter instrument (W0173), reproduced from \cite{rabault2020development}. The instruments were deployed close to each
other, but drifted slightly with time. Still, agreement within 5\% is obtained for the significant wave height estimate (Hs, top), and the location
and height of the wave spectrum peak (bottom). Slight differences are observed regarding the high frequency tail of the spectra. This is due to the increased
noise of the GPS-based measurement methodology of the Sofar Spotter in quiet conditions corresponding to waves in ice, compared with the accelerometer-based
measurement methodology of the v2018 in the same conditions.}
\label{fig:v2018_validation_sofar}
\end{figure}

\begin{figure}[ht]
\centering
\includegraphics[width=0.60\linewidth]{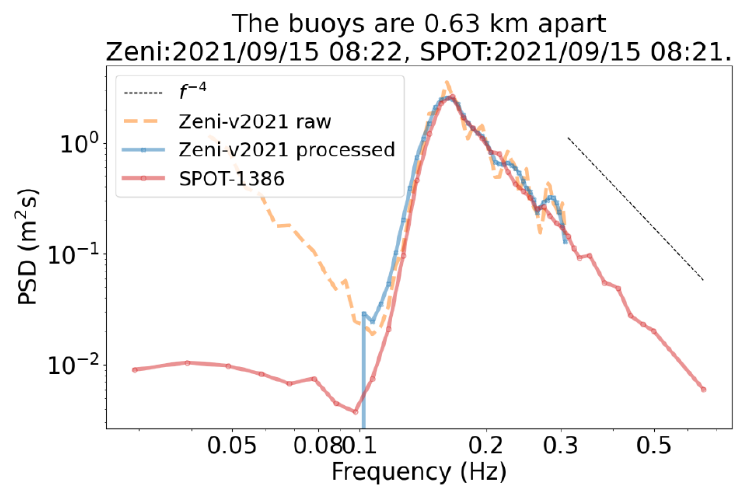}
\caption{Validation of the instrument v2021 (Zeni-v2021 processed) against the Sofar Spotter (SPOT-1386), in open water, reproduced from \cite{rabault2022openmetbuoy}. The instruments were deployed close to each
other, but drifted significantly with time. Agreement within typically 5\% is obtained on most of the spectrum. The increasing, spurious noise at low frequencies
observed in the Zeni-v2021, is a well known phenomena for IMU-based measurements of waves in ice in the open ocean \cite{nose2018predictability}, which can be filtered out. This phenomenon
is not expected to be present when performing measurements of waves in the sea ice.}
\label{fig:v2021_validation_sofar}
\end{figure}

\begin{figure}[ht]
\centering
\includegraphics[width=\linewidth]{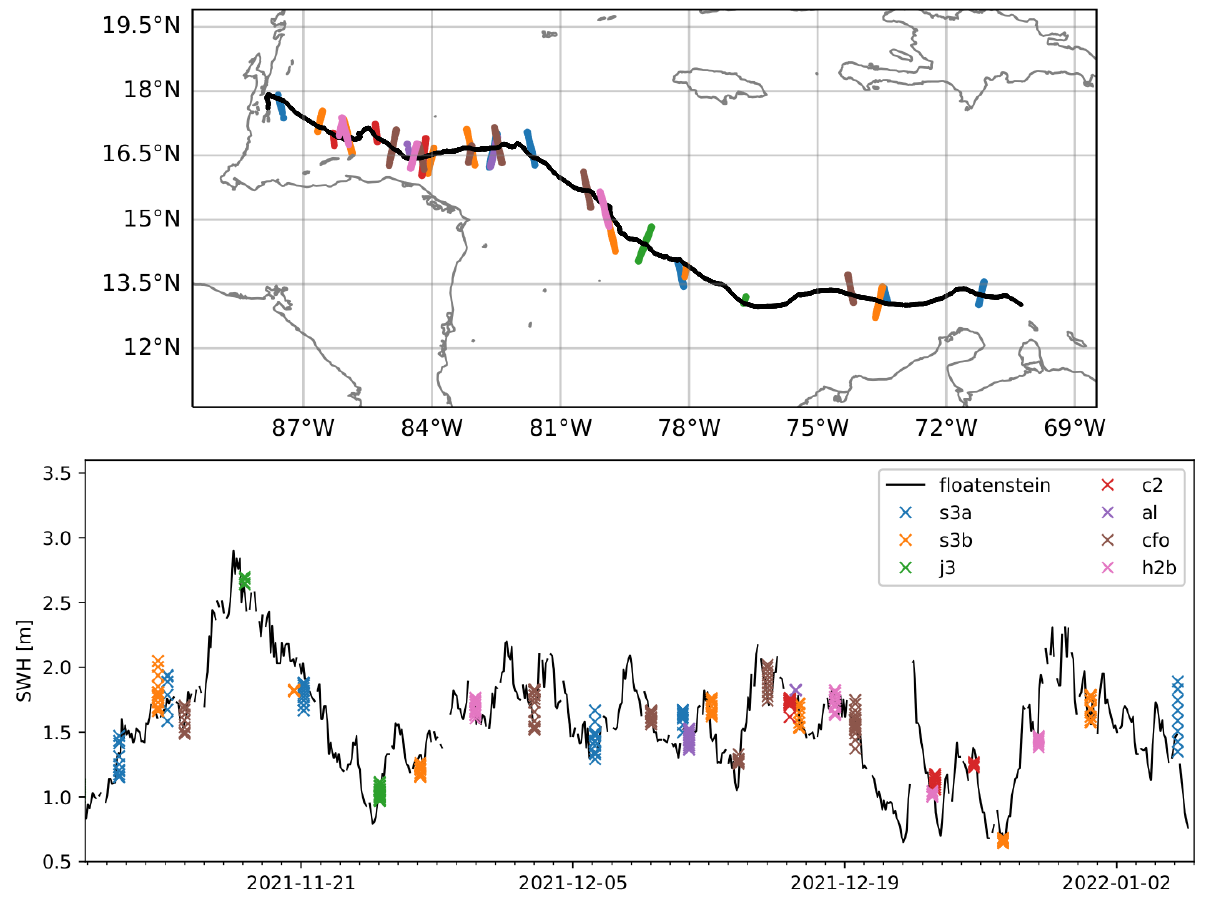}
\caption{Validation of the instrument v2021 against satellite measurements, reproduced from \cite{rabault2022openmetbuoy}. Top: drift trajectory of a freely drifting, open water version of the
instrument v2021 (black), and illustration of satellite swaths intersecting the trajectory over the drift period (colored point clouds).
Bottom: comparison of the significant wave height (SWH) reported from the v2021 (nicknamed "floatenstein"), with satellite measurements from a variety of satellites. SWH comparison
between the v2021 and the satellite measurements are in agreement, with the measurements from the v2021 always falling within the spread for each satellite measurement swath. While this is obtained from
a deployment performed in the Caribbeans, this is a validation of the correct functioning of the wave measurement routines.}
\label{fig:v2021_validation_satellite}
\end{figure}

\begin{figure}[ht]
\centering
\includegraphics[width=0.50\linewidth]{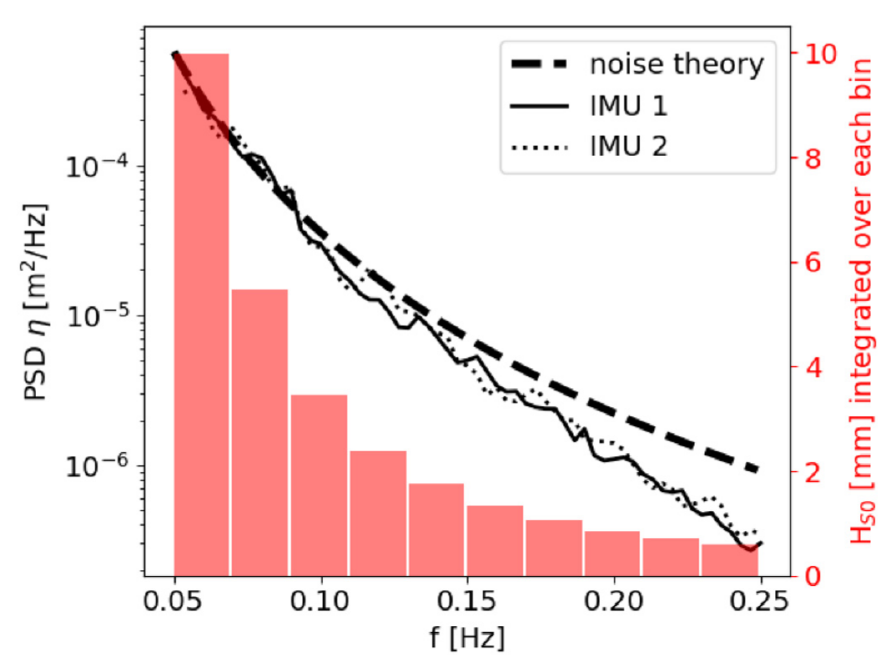}
\caption{Illustration of the noise threshold at rest for the instrument v2018, reproduced from \cite{rabault2020open}. This characterizes the typical noise level of the combination of the VN100
IMU and the wave processing algorithm used to generate the spectra transmitted over Iridium.}
\label{fig:v2018_noise}
\end{figure}

\begin{figure}[ht]
\centering
\includegraphics[width=0.60\linewidth]{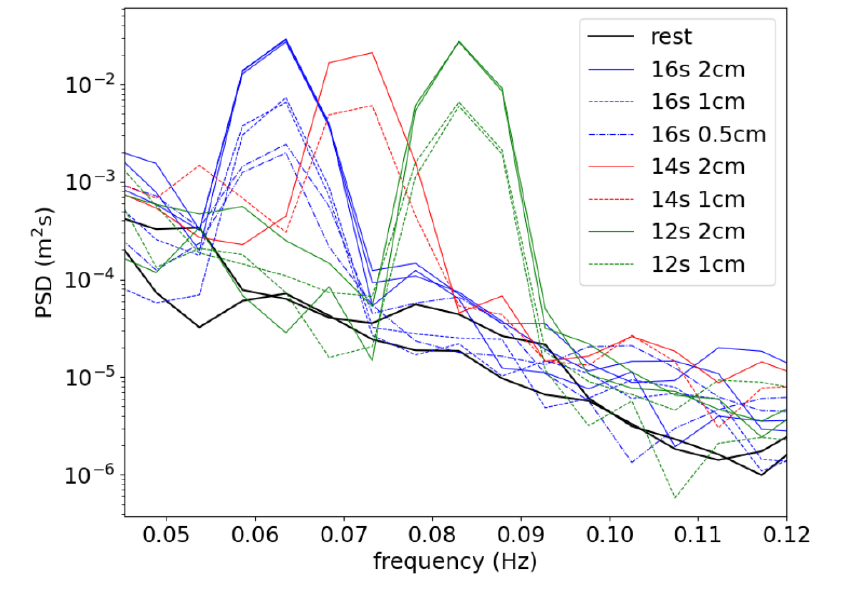}
\caption{Illustration of the noise threshold at rest and under a variety of wave motions produced artificially in the laboratory for the instrument v2021, reproduced from \cite{rabault2022openmetbuoy}. This characterizes the typical noise level resulting from the combination of
raw data measurements by the accelerometer and gyroscope, and the processing algorithms used.}
\label{fig:v2021_noise}
\end{figure}

Results that come from in-situ processing of IMU or GPS data, such as the wave spectra and statistics reported by the
Sofar Spotter and the instruments v2018 and v2021, have been previously validated in details, and the reports and validation details are available
in the literature: the Sofar Spotter is a commercial instrument that was validated before being
released for sale \cite{PerformanceCharacteristicsofSpotteraNewlyDevelopedRealTimeWaveMeasurementBuoy}, the v2018 has been validated and used in scientific papers multiple times \cite{rabault2020open, voermans2020experimental, voermans2021wave}, and the v2021 has
been recently validated against both commercial buoys and satellite data \cite{rabault2022openmetbuoy}. Validation
campaigns for both the v2018 and the v2021 indicated agreement to either within 5\%, or within one standard deviation, of commercial instrument or other measurement methodology. While
more details are available in the corresponding papers, we reproduce the main validation figures against established commercial instruments in Figs. \ref{fig:v2018_validation_seabird},\ref{fig:v2018_validation_sofar} (v2018) and Figs. \ref{fig:v2021_validation_sofar}, \ref{fig:v2021_validation_satellite} (v2021), for the sake of completeness. In
addition, since the v2018 and the v2021 have much higher levels of sensitivity compared with GPS-based buoys or satellite measurements against which we validated them, tests were performed in the laboratory, in controlled conditions,
in order to estimate the noise background of the whole system (i.e., including the noise of the IMU itself, and the effect of the processing algorithms used, as described in the Methodology section). As visible in these tests,
which main findings are reproduced in Fig. \ref{fig:v2018_noise} (v2018), and Fig. \ref{fig:v2021_noise} (v2021), the instruments are able to measure waves with amplitude down to typically a few millimeters. Test results are in agreement with the accelerometer datasheet (which describe the expected noise background intensity), and the formula for conversion from acceleration to elevation spectrum (Eqn. \ref{psd_to_spectrum}, which describes the expected spectral shape of the noise), where the noise backgrounds of the v2018 and the v2021
follow the expected decay curve as frequency increases (i.e., accelerometer-based instruments are more sensitive for wave amplitude as wave frequency increases). The typical noise threshold
for a single data bin for the instrument v2018 is about 10mm for 20s waves, and 0.3mm for 4s waves, as visible in Fig. 3 of \cite{rabault2020open}, reproduced here as Fig. \ref{fig:v2018_noise}. The typical noise
threshold for the v2021 is even lower, as visible in the Fig. 5 of \cite{rabault2022openmetbuoy}, reproduced here as Fig. \ref{fig:v2021_noise}, as a 16s wave of amplitude 5mm corresponds to a signal to noise ratio (SNR) of around 10. 

These validations, both against other field instruments, against satellite measurements, and in controlled laboratory conditions,
give us confidence in the accuracy and reliability of our instruments.

\section*{Usage Notes}

Two kinds of data processing levels are provided to the user:

\begin{itemize}
  \item High-level data that are ready-for-use are provided as netCDF-CF files, following the
  best practices in use in the geoscience community and the FAIR principles \cite{wilkinson2016fair}. These are available on a specific folder of the THREDDS server
  of the Norwegian Meteorological Institute as part of the Arctic Data Center repository \url{https://doi.org/10.21343/AZKY-0X44}\cite{thredds_data}. Any netCDF package, in python or other language, can be used to access the data.
  The use of NetCDF4-CF files ensures long time compatibility, as this is the standard used for
  archiving meteorological data all over the world, and the tooling is mature and stable.
  
  \item In addition, the full raw data, either SD card records when the instruments could be
  recovered, or raw iridium transmissions, are available at the Github repository that supports
  the data release \url{https://github.com/jerabaul29/data_release_sea_ice_drift_waves_in_ice_marginal_ice_zone_2022} \cite{main_gh_repo}. Custom
  software is provided (either on the same repository, or on the repositories where the source code of the
  instruments and post processing scripts are available) to decode these raw data. The fact that both the
  firmware of the instruments (hence, the binary protocol used), and the binary decoders, are open source,
  guarantees long term accessibility of the data, as well as full insight into the technical details associated.
\end{itemize}

No additional processing is needed to use either the GPS or the wave data, and the data are ready to
use as-is in Python or similar.

A proper use of the presented data for the analysis of wave damping will also critically depend on the availability of high-quality sea-ice concentration, thickness, and floe size distribution maps, as well as the wave frequency spectra in the open ocean adjacent to the sea-ice edge. While providing these data is, in general, outside of our scope, as there are a number of concurrent, different possible sources
and models that produce such data, and these data are generated independently of the field measurements and rely on entirely different techniques, codes, and operational products, we highlight
a few different sources that could be of interest to the reader.

Sea-ice concentration products are available from passive microwave sensors with a footprint of around O(20km) \cite{kern2019satellite}. New multi-sensor products are becoming available, combining  sensor information of passive and active microwave and Synthetic Aperture Radar (SAR) sensors \cite{dierking2013sea, 2017EGUGA1919037D, zakhvatkina2019satellite}. These products reach resolutions of O(1 to 10km). In addition, manual sea-ice charts can be used which cover scales down to O(100m) and are available on a daily basis from the national ice services (for example, the Norwegian Meteorological Institute Ice Service charts, \url{https://cryo.met.no/en/latest-ice-charts}, or the similar Danish Meteorological Institute Icecharts, \url{http://ocean.dmi.dk/arctic/icecharts.uk.php}). Various pan Arctic sea-ice thickness satellite products are available with resolutions of O(25km) and are considered reliable in the thickness range from 0.5 to 4 m \cite{sallila2019assessment}. Approaches to produce satellite based information on the sea-ice floe size distribution are emerging but to our knowledge no operational products are available yet \cite{horvat2019estimating}. Some national ice services (e.g. the Danish Sea Ice Service) include information on floe sizes as part of the manual ice charts \cite{oceanographicsigrid}. In addition to the sea-ice information, wave information on the wave properties adjacent to the sea-ice edge are needed. A number of wave hindcasts and reanalysis are available, however, often the energy frequency spectra are not disseminated (e.g. \cite{Carrasco2022}); we are currently providing this feedback to the modeling community, and hope that more detailed information from numerical models will be available in the future.

\section*{Code availability}

The firmwares of the instruments v2018 and v2021, as well as the binary data decoder scripts, are fully available on the corresponding github repositories
previously mentioned. The scripts to plot the data from the netCDF files are available on the main Github repository previously mentioned, \url{https://github.com/jerabaul29/data_release_sea_ice_drift_waves_in_ice_marginal_ice_zone_2022},
together with the raw data.
All code is developed in modern python (version 3.8 or higher), or C++, or Matlab, unless specified otherwise. The netCDF datafiles
are following the netCDF4 standard, with CF attributes conventions. We also provide a mirror of the netCDF data files on the THREDDS server of the Norwegian Meteorological Institute in the context of the Arctic Data Center repository, at the following address: \url{https://doi.org/10.21343/AZKY-0X44}.

We will offer reasonable support regarding the data and its use through the Issues tracker of the data repository at \url{https://github.com/jerabaul29/data_release_sea_ice_drift_waves_in_ice_marginal_ice_zone_2022}, and we invite readers in need of specific help to take contact with us there. In addition,
we plan on releasing extensions to this dataset periodically as more data are collected. We invite scientists who own similar data and are willing to release these as
open source materials to take contact with us so that they can get involved in the next data release we will perform.

In addition, we discovered, in the context of the present work, that there are already some openly available data about sea ice drift and waves in ice (to cite but a few, \cite{SIPEXII, long_term_meas_Thomson, wave_driven_flow_data}); however, these are scattered across the internet, and may be difficult to find. Therefore, in addition to the data release intrinsic to this dataset, we have started to maintain an index of similar open data at \url{https://github.com/jerabaul29/meta_overview_sea_ice_available_data}. We invite the reader aware of additional open datasets to notify us so that these can be added to our index, which we will keep extending in the future. We hope that these data, together with a variety of datasets that have been recently gathered \cite{schulz2022full, lei2021seasonal}, will be a significant contribution towards building large, well sampled datasets of in situ observations of the MIZ and sea ice dynamics.

\bibliography{sample}

\section*{Acknowledgements}

This work was partly funded through the following projects and funding agencies:

\begin{itemize}

    \item the Norwegian Research Council, the
    Norwegian Polar Institute, and the Norwegian Meteorological Institute through the following projects:
    Dynamics Of Floating Ice (DOFI, grant number 280625), Arven Etter Nansen (AeN, 276730),
    Machine Ocean (303411), FOCUS (301450), CIRFA (237906).

    \item JV and AB were supported by the Australian Antarctic Science Program under 
    Project AAS4593. AB acknowledges support from the US Office of Naval Research Grant
    Number N62909-20-1-2080.

    \item AM and IT were supported by the Research Council of Norway through the Centre for Sustainable Arctic Marine and Coastal Technology (SAMCoT), the Arctic Offshore and Coastal Engineering in a Changing Climate (AOCEC) project of the Programme for International Partnerships for Excellent Education, Research, and Innovation (IntPart, Project number 274951), and the Dynamics of Floating Ice (DOFI) project of the Large-scale Programme for Petroleum Research (Petromaks2).

    \item YK was supported by Lundin Energy Norway (grant 802144).

\end{itemize}

We want to thank the crews of the R/V Kronprins Haakon and the MS Polarsyssel for their invaluable help during fieldwork. IT and AM thank the C-CORE for in-kind support of this work and Ryan Crawford and Erik Veitch for their assistance with the fieldwork. We also want to thank Jim Thomson for his support of our idea to create a github data index, and pointing us to the data he has been involved in releasing openly on data repositories.

\section*{Author contributions statement}

JR and MM were the main initiators of the manuscript and designed the data collection and its scope. JR and PT put the data in the common
format, and prepared the technical aspects and data management aspects associated with the data release.

JR designed the instruments v2018 (with help from GS and OG) and the instrument v2021 (with help from MM, ØB, GH). ARM designed
the IWR. JR, MM, JB, TN, MJ, ØB, GH, TK, VdA, CT, built instruments used in specific deployments.

MM, AM, MB, TW, ØB, MJ, JR, AJ, YK, AB, KHC, designed, obtained funding for, organized, and lead, individual deployments and field campaigns.

JR, MM, JV, DB, IT, AM, MB, TN, TW, MJ, ØB, GS, LRH, MJ, ARM, AJ, YK, MK, GH, TK, VdA, CT, CPQ, KF, designed, performed, and collected
the data from individual deployments reported in the manuscript.

JR, MM wrote the main parts of the manuscript. JR, JV, DB, IT, AM, MB, TN wrote specific parts of the manuscripts corresponding
to some individual deployments. All authors reviewed the manuscript and participated in iterating the manuscript.

\section*{Competing interests}

The authors declare no competing interests nor ethical concerns.

\section*{Figures \& Tables}

\begin{table}[ht]
\centering
\begin{tabular}{|c|c|c|c|}
\hline
Tracker ID & deployment time (UTC) & floe dimensions (m) & floe surface area (m$^2$) \\
\hline
\hline
4610 & April 24 2017, 08:54 & 30 x 43 & 934 \\

9630 & April 25 2017, 07:59 & 10 x 25 & 250 \\
2620 & April 25 2017, 08:05 & 10 x 15 & 150 \\
3620 & April 25 2017, 08:10 & 5 x 8 & 40 \\
0630 & April 25 2017, 08:15 & 20 x 30 & 600 \\
2470 & April 26 2017, 08:20 & 10 x 25 & 250 \\
\hline
\end{tabular}
\caption{\label{tab:ice_tracker_deployment_2017}Details around the deployment conditions
for the 8 ice trackers, Svalbard Banks, 2017-04.}
\end{table}

\begin{table}[ht]
\centering
\begin{tabular}{|c|c|c|c|}
\hline
Deployment time & location & ice conditions & number \& kind of instrument \\
\hline
\hline
2017-04 & Arctic, Barents Sea, 76.4N 22.5E & drift ice: 8/10 to 0/10 & GPS drifter: 8 \\
\hline
2018-03a & Arctic, East Greenland Sea, 73.5N 15.5E & drift ice: 6/10 to 10/10 & GPS drifter: 5 \\
\hline
2018-03b & Arctic, Beaufort Sea, 72.3N 148.4W & pack ice: 8/10 to 10/10 & IWR: 2 \\
\hline
2018-04 & Arctic, Barents Sea, 75.3N 19.5E & drift ice: 8/10 to 0/10 & GPS drifter: 1 \\
\hline
2018-09 & Arctic, Barents Sea, 82N 20E & MIZ: 1/10 to 10/10 & v2018: 4 \\
\hline
2020-01 & Antarctic, outside Davis station, 69S 76E & landfast ice (breakup) & v2018: 2 +
Sofar Spotter: 2 \\
\hline
2020-03a & Arctic, Grønfjorden, Svalbard, 78N 14E & landfast ice (intact) & v2018: 3 \\
\hline
2020-03b & Arctic, Beaufort Sea, 71.2N 141.5W & pack ice: 8/10 to 10/10 & IWR: 2 \\
\hline
2020-07 & Arctic, Yermak Plateau, Barents Sea, 82N 15E & MIZ: 3/10 to 10/10 & v2018: 6 \\
\hline
2020-11 & Antarctic, outside Casey station, 66S 110E & landfast ice (intact) & v2018: 2 \\
\hline
2021-02 & Arctic, Barents Sea, east Svalbard, 77N 30E & MIZ: 5/10 to 10/10 & v2018: 6 +
v2021: 11 (6 with waves) \\
\hline
2021-03 & Arctic, Beaufort Sea, 71.5N 148WE & pack ice: 7/10 to 10/10 & IWR: 3 \\
\hline
2021-04 & Arctic, Utqiagvik, 71.3N 156.6W & landfast ice & IWR: 6 \\
\hline
2021-09 & Arctic, Laptev Sea, 82N 118E & MIZ: 1/10 to 10/10 & v2021: 1 +
Sofar Spotter: 1 \\
\hline
2022-03 & Arctic, East Greenland sea, 70N 20E & MIZ: 2/10 to 10/10 & v2021: 2 + commercial beacon: 5\\
\hline
\hline
total nbr tracks & & & total: 72; with waves: 48 \\
\hline
\end{tabular}
\caption{\label{tab:deployments}Overview of the deployments, their locations and time spans, the
sea ice conditions, and the
kind and number of instruments deployed.}
\end{table}

\end{document}